\documentclass[twocolumn,showpacs,preprintnumbers,amsmath,amssymb]{revtex4}

\usepackage{graphicx}
\usepackage{dcolumn}
\usepackage{bm}

\begin{document}


\title{Aging to Equilibrium Dynamics of SiO$_2$}

\author{K. Vollmayr-Lee}
 \email{kvollmay@bucknell.edu}
\author{J.\ A. Roman}%
\affiliation{%
Department of Physics and Astronomy, Bucknell University,
      Lewisburg, Pennsylvania 17837, USA
}%

\author{J. Horbach}
\affiliation{
Institut f\"ur Materialphysik im Weltraum,
Deutsches Zentrum f\"ur Luft- und Raumfahrt (DLR),
Linder H\"ohe, 51147 K\"oln, Germany
}%

\date{May 3, 2010}

\begin{abstract}
Molecular dynamics computer simulations are used to study the aging
dynamics of SiO$_2$ (modeled by the BKS model).  Starting from
fully equilibrated configurations at high temperatures $T_{\rm
i} \in \{5000$\,K$, 3760$\,K\}, the system is quenched to lower
temperatures $T_{\rm f} \in \{ 2500$\,K, 2750\,K, 3000\,K, 3250\,K$\}$
and observed after a waiting time $t_{\rm w}$.  Since the simulation
runs are long enough to reach equilibrium at $T_{\rm f}$, we are able to
study the transition from out-of-equilibrium to equilibrium dynamics.
We present results for the partial structure factors, for the generalized
incoherent intermediate scattering function $C_q(t_{\rm w}, t_{\rm w}+t)$,
and for the mean square displacement $\Delta r^2(t_{\rm w}, t_{\rm w}+t)$.
We conclude that there are three different $t_{\rm w}$ regions: (I) At
very short waiting times, $C_q(t_{\rm w}, t_{\rm w}+t)$ decays very fast
without forming a plateau. Similarly $\Delta r^2(t_{\rm w}, t_{\rm w}+t)$
increases without forming a plateau.  (II) With increasing $t_{\rm w}$ a
plateau develops in $C_q(t_{\rm w}, t_{\rm w}+t)$ and $\Delta r^2(t_{\rm
w}, t_{\rm w}+t)$. For intermediate waiting times the plateau height is
independent of $t_{\rm w}$ and $T_{\rm i}$.  Time superposition applies,
i.e. $C_q=C_q(t/t_{\rm r}^{\rm Cq})$ where 
$t_{\rm r}^{\rm Cq}=t_{\rm r}^{\rm Cq}(t_{\rm w})$
is a waiting time dependent decay time. Furthermore 
$C_q=C(q,t_{\rm w},t_{\rm w}+t)$
scales as $C_q=C(q,z(t_{\rm w},t)$ where $z$ is a function of $t_{\rm w}$ and
$t$ only, i.e.~independent of $q$.  (III) At large
$t_{\rm w}$ the system reaches equilibrium, i.e.~$C_q(t_{\rm w},
t_{\rm w}+t)$ and $\Delta r^2(t_{\rm w}, t_{\rm w}+t)$ are independent
of $t_{\rm w}$ and $T_{\rm i}$. For $C_q(t_{\rm w}, t_{\rm w}+t)$  we find 
that the time superposition of intermediate waiting times (II) includes the 
equilibrium curve (III).
\end{abstract}
%
\maketitle
\section{Introduction}
When a glass-forming liquid is quenched from an equilibrium state at
a high temperature $T_{\rm i}$ to a non-equilibrium state at a lower
temperature $T_{\rm f}$, ``aging processes'' set in.
Provided that crystallization plays no role at
$T_{\rm f}$ (e.g.~due to very low crystal nucleation rates), the transition
to a final (metastable) equilibrium state occurs on a time scale that
corresponds to the typical equilibrium relaxation time $\tau_{\rm eq}$
of the (supercooled) liquid at $T_{\rm f}$.  
The dynamics of the system depends on the waiting time $t_{\rm w}$ which 
is the time elapsed after the temperature quench.
If $\tau_{\rm eq}$ exceeds
the waiting time $t_{\rm w}$ then the system
is observed in a transient non-equilibrium state which corresponds to a glass 
for $t_{\rm w} \ll \tau_{\rm eq}$.
During the aging process, i.e.~for $t_{\rm w} < \tau_{\rm eq}$,
thermodynamic properties such as volume and energy are changing and
time translation invariance does not hold: correlation functions at time
$t_{\rm w}+t$ and the time origin at $t_{\rm w}$ do depend
not only on the time difference $t$ but also on the waiting time $t_{\rm w}$.

Recently this aging process has been investigated extensively
with experiments \cite{cipelletti,lynch08,cianci,courtland,pham,habdas}, 
theoretically \cite{cugliandolo93,MFCq51,MFCq52}  
and with computer simulations. For a more complete summary of previous 
results we refer the reader to the references  
\cite{leshouches2002,glassbook} and references therein.
Computer simulation studies most similar to the work presented here 
are on attractive colloidal systems 
\cite{foffi04,foffi05PRL,foffi05JCP,puertas07}, 
on the Kob-Andersen Lennard-Jones (KALJ) mixture
\cite{KobBarratPRL78,KobBarratCq1Cq2,barrat99,kob00,sciortino01,saika04,parsaeian08,parsaeian09,mossa02,berthier07}, and silica (SiO$_2$) \cite{WahlenRieger,scala03,berthier07,parsaeian10}. 
In the case of silica the interpretation of the results is less clear
than for the KALJ mixture; e.g.~different findings \cite{scala03,berthier07} have been reported
on the violation of the fluctuation-dissipation regime 
during aging \cite{cugliandolo93,MFCq51,MFCq52}. 
Thus, it remains open whether silica, as the
prototype of a glass-forming system forming a tetrahedral network structure,
exhibits a different aging dynamics than, e.g.~the KALJ model, where
the structure is similar to that of a closed-packed hard-sphere structure.

Recent simulation studies on amorphous silica 
\cite{vollmayr96_2,WahlenRieger,BKSTc,parsaeian10,JuergenFSFs,berthier07,vink03,poole01,poole04,scala03,saksaengwijit2004,reinisch2005,saksaengwijit2006,saksaengwijit2007,scheidler2001}
have widely used the
BKS potential \cite{beest_90} to model the interactions between the
atoms. Although it is a simple pair potential, it reproduces
various static and dynamic properties of amorphous silica very well. For
BKS silica, the self-diffusion constants $D_{\alpha}$ ($\alpha={\rm
Si,O}$) show two different temperature regimes: At high temperatures,
$D_{\alpha}$ decays according to a power law, as predicted by the mode
coupling theory (MCT) of the glass transition (note, however, that also
other interpretations have been assigned to this high temperature regime).
At low temperature, $D_{\alpha}$ as well as the shear viscosity $\eta$
exhibit an Arrhenius behavior with an activation energy of the order
of 5\,eV, in agreement with experiment (see \cite{BKSTc} and references
therein).  The temperature at which the crossover between both regimes
occurs is at $T_c\approx3300$\,K, corresponding to the critical MCT
temperature of BKS silica.  Previous studies of the aging dynamics of BKS
silica \cite{WahlenRieger,scala03,berthier07} were performed in two steps.
First, the system was fully equilibrated at a temperature $T_{\rm i}>T_c$. Then,
the system was quenched to a low temperature $T_{\rm f}<T_c$, followed by 
the production runs. Wahlen and Rieger \cite{WahlenRieger} 
analyze time-dependent correlation functions at different waiting
times $t_{\rm w}$ and Berthier \cite{berthier07} and 
Scala et al.\ \cite{scala03} study the generalized 
fluctuation dissipation relation and the energy landscape \cite{scala03}. 
All three studies 
\cite{WahlenRieger,scala03,berthier07} investigate 
the early stages of the aging dynamics, i.e.~the dynamics
was explored on time scales that were much smaller than the equilibrium
relaxation time $\tau_{\rm eq}$ at the temperature $T_{\rm f}$.

In this work, we also consider quenches in BKS silica from a high
temperature $T_{\rm i}$ to a low temperature $T_{\rm f}$. Different from
previous simulation studies, we aim at elucidating the full transient
dynamics at $T_{\rm f}$ from the initial state at $t_{\rm w}=0$ to
equilibrium.  To this end, temperatures $T_{\rm f}$ are chosen such that
the system can be fully equilibrated on the typical time span of the
MD simulation.  Note that the considered 
temperatures $T_{\rm f} \in \{ 2500$\,K, 2750\,K, 3000\,K, 3250\,K$\}$
are below the critical MCT temperatures $T_{\rm c}$. Thus, we have access
to the full aging dynamics in the experimentally relevant Arrhenius
temperature regime that we have mentioned above. 

The analysis of time-dependent density correlation functions $C_q(t_{\rm
w}, t_{\rm w}+t)$ (with $q$ the wavenumber) and the mean square displacement
$\Delta r^2(t_{\rm w}, t_{\rm w}+t)$ reveal three different
regimes of waiting times $t_{\rm w}$: In the case of $C_q(t_{\rm
w}, t_{\rm w}+t)$ (and similarly for $\Delta r^2(t_{\rm w}, t_{\rm w}+t)$)
at early $t_{\rm w}$, a rapid decay
to zero is seen, without forming a plateau at intermediate times.  Then,
for larger values of $t_{\rm w}$ a plateau is formed. The height of this
plateau grows with waiting time and becomes more pronounced, before in the
final regime the plateau height is independent of $t_{\rm w}$ and $T_{\rm
i}$. In the latter regime, time superposition holds, i.e.~by scaling time
with a decay time $t_{\rm r}^{\rm Cq}$ the $C_q$ for the 
different values of $t_{\rm w}$ fall
onto a master curve at a given wavenumber $q$.  This behavior is very
similar to that found for the KALJ mixture. However, it is different
from the behavior predicted by mean-field spin glass models and the
activated dynamics scaling, as proposed by Wahlen and Rieger.  Thus,
these results suggest that the aging dynamics in silica, the prototype
of a glass-former with a tetrahedral network structure, is very similar
to that of simple glass-formers with a closed-packed hard-sphere-like
structure. We find a difference between the KALJ mixture and SiO$_2$,
however, in the parametric plot of $C_q'(C_q)$. For SiO$_2$ $C_q'(C_q)$
shows a data collapse for different sufficiently large $t_{\rm w}$ and thus
$C_q=C(q,z(t_{\rm w},t))$ whereas this data collapse 
does not hold as well in the case of the KALJ mixture.

The rest of the paper is organized as follows: In the next Sec. we give
the details of the BKS potential and the simulation. Then, we present the
results in Sec. III, before we summarize in Sec. IV. Appendix A describes
the implementation of the Nos\'e-Hoover thermostat used in our simulation.

\section{Model and Details of The Simulation}
The interactions between the particles are modeled by the BKS
potential \cite{beest_90} which has been used frequently and has
proven to be reliable for the study of the dynamics of amorphous silica
\cite{vollmayr96_2,WahlenRieger,BKSTc,parsaeian10,JuergenFSFs,berthier07,vink03,poole01,poole04,scala03,saksaengwijit2004,reinisch2005,saksaengwijit2006,saksaengwijit2007,scheidler2001}
The functional form of the BKS potential is
given by a sum of a Coulomb term, an exponential and a van der Waals
term. Thus the potential between particles $i$ and $j$, a distance
$r_{ij}$ apart, is given by
\begin{equation}
\phi(r_{ij})=\frac{q_i q_j e^2}{r_{ij}}+A_{ij}e^{-B_{ij}r_{ij}}-
\frac{C_{ij}}{r_{ij}^6}\quad ,
\label{eq1}
\end{equation}
where $e$ is the charge of an electron and the constants $A_{ij}$,
$B_{ij}$ and $C_{ij}$ are $A_{\mbox{\footnotesize SiSi}}=0.0$\,eV,
$A_{\mbox{\footnotesize SiO}}=18003.7572$\,eV, $A_{\mbox{\footnotesize
OO}}=1388.7730$\,eV, $B_{\mbox{\footnotesize SiSi}}=0.0$\,\AA$^{-1}$,
$B_{\mbox{\footnotesize SiO}}=4.87318$\,\AA$^{-1}$,
$B_{\mbox{\footnotesize OO}}=2.76000$\,\AA$^{-1}$, $C_{\mbox{\footnotesize
SiSi}}=0.0$\,eV\AA$^{-6}$, $C_{\mbox{\footnotesize
SiO}}=133.5381$\,eV\AA$^{-6}$ and $C_{\mbox{\footnotesize
OO}}=175.0000$\,eV\AA$^{-6}$~\cite{beest_90}.  The partial charges
$q_{i}$ are $q_{\mbox{\footnotesize Si}}=2.4$ and $q_{\mbox{\footnotesize
O}}=-1.2$ and $e^2$ is given by $1602.19/(4\pi 8.8542)$\,eV\AA.

The Coulombic part of the interaction was computed by using the Ewald
method~\cite{KiefferAngell,allen90} with a constant $\alpha L=6.3452$,
where $L$ is the size of the cubic box, and by using all $q$-vectors
with $|q|\leq 6\cdot 2\pi /L$.
\footnote{The erfc was approximated with a polynomial of fifth 
order \cite{erfcpol}.}.
We ensure that the Ewald term in real space is also
differentiable at the cutoff by smoothing similarly to Eq.~(3)
in \cite{BKSPotDetails:Pfleiderer} with $r_{\rm c}=8$\,\AA\
and $d=0.05$\,\AA$^2$.  To increase computation speed the
non-Coulombic contribution to the potential was truncated, smoothed
and shifted at a distance of 5.5\,\AA. Note that this truncation
is not negligible since it affects the pressure of the system.
In Ref.~\cite{BKSPotDetails:Pfleiderer} further slight variations on
the potential are described in detail  
\footnote{Please note that we chose all parameters as described in
\cite{BKSPotDetails:Pfleiderer} with the only exception of $a_{2,{\rm
O}}=895.11679$\,eV/\AA  (instead of $a_{2,{\rm O}}=90.38499$ eV/\AA).}.
In order to minimize surface effects periodic boundary conditions were
used. The masses of the Si and O atoms were 28.086\,u and 15.9994\,u,
respectively. The number of particles was 336, of which 112 were silica
atoms and 224 were oxygen atoms.  For all simulation runs the size of
the cubic box was fixed to $L=16.920468$\,\AA\ which corresponds to a
density of $\rho=2.323$\,g/cm$^3$, a value that is very close to the
one of real silica glass, $\rho=2.2$\,g/cm$^3$ \cite{Brueckner}.

\begin{figure}
\includegraphics[width=3.4in]{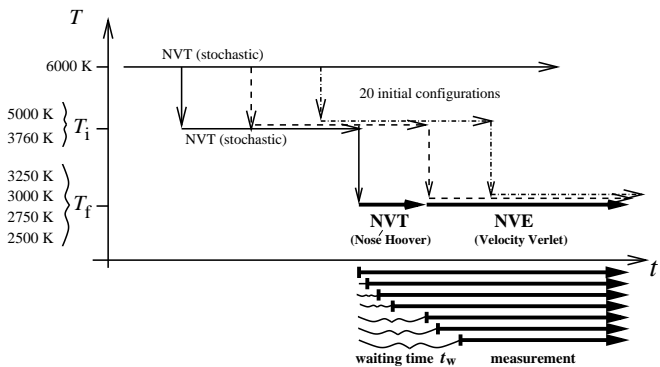}
\caption{\sf
Schematic sketch of the protocol for the simulation runs.
20 independent initial configurations are obtained via one long simulation
run at $6000$ K. For each independent configuration the system is
quenched instantaneously and then 
fully equilibrated at 
temperatures $T_{\rm i}=3760$\,K and 5000\,K, followed by 
instantaneous quenches
to the temperatures $T_{\rm f}=3250$\,K, $3000$\,K, $2750$\,K,
and $2500$\,K. At each $T_{\rm f}$, time-dependent correlation functions
are determined for different waiting times $t_{\rm w}$.  Temperature is
kept constant at $T_{\rm f}$ by coupling the system to a Nos\'e-Hoover
thermostat.  The thermostat is switched off after 0.33\,ns, followed by
the continuation of the simulations in the microcanonical ensemble 
for $33$ ns.}
\label{coolsketch}
\end{figure}
\begin{figure}
\includegraphics[width=3.4in]{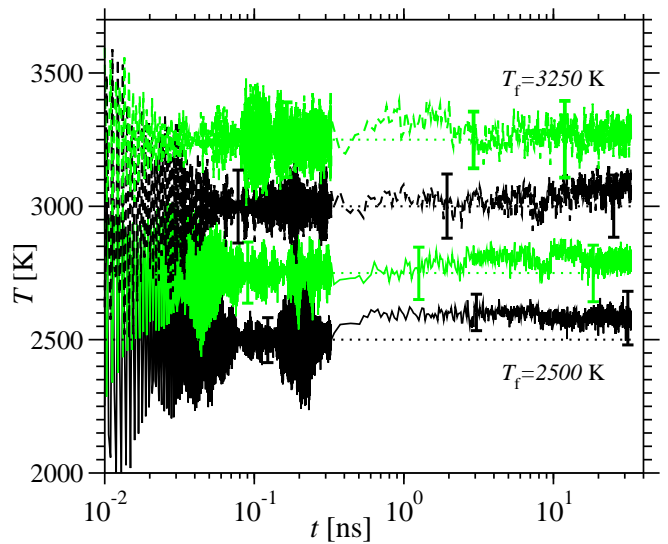}
\caption{\sf  
(Color online) Temperature $T=\frac{2 {\overline{E}}_{\rm kin}}{3 N}$ 
as a function of time $t$ for 
$T_{\rm i}=5000\,{\rm K}$, $T_{\rm f}=2500\,{\rm K},
2750\,{\rm K}, 3000\,{\rm K}, 3250\,{\rm K}$ shown in each case for 
the 11th independent simulation run. ${\overline{E}}_{\rm kin}$ includes 
a time average and error bars indicate the corresponding fluctuations.
}
\label{tempoftav_Tallc11}
\end{figure}

We investigated the aging dynamics for systems which were quenched
from a high temperature $T_{\rm i}$ to a low temperature $T_{\rm f}$.
To increase the statistics, for each $(T_{\rm i}, T_{\rm f})$
20 independent simulation runs were performed. To obtain 20 independent
configurations we carried out molecular dynamics (MD) simulations
using the velocity Verlet algorithm with a time step of $1.6$\,fs at
6000\,K. The temperature was kept constant at $6000$ K 
with a stochastic heat bath
by replacing the velocities of all particles by new velocities drawn from
the corresponding Boltzmann distribution every 150 time steps. Independent
configurations were at least $3.27$ ns 
apart.  Each of these
configurations undergoes the following sequence of simulation runs (see
also Fig.~\ref{coolsketch}).  After fully equilibrating the samples at
the initial temperatures $T_{\rm i} = 5000$\,K 
(for $16.35$ ns) 
and $T_{\rm i} = 3760$\,K 
(for $32.7$ ns), 
the system was quenched instantaneously to $T_{\rm f} \in \{2500\,{\rm K},
2750\,{\rm K}, 3000\,{\rm K}, 3250\,{\rm K} \}$. To disturb the dynamics
minimally, we used a Nos\'e-Hoover thermostat \cite{hoover,branka}
instead of a stochastic heat bath to keep the temperature at $T_{\rm f}$ 
constant.
A velocity Verlet algorithm was used to integrate the Nos\'e-Hoover
equations of motion (see Appendix \ref{NoseHooverEqu}) with a time step
of 1.02\,fs. After $0.33$ ns 
the Nos\'e-Hoover thermostat was
switched off and the simulation was continued in the NVE ensemble for
$33$ ns 
using a time step of $1.6$\,fs.
Whereas previous simulations used instead the NVT ensemble for the 
whole simulation run, we chose to switch to the NVE ensemble to 
minimize any influence on the dynamics due to the chosen 
heat bath algorithm. 
For the comparison with previous simulations and to check 
for the lack of a temperature drift, we show in 
Fig.~\ref{tempoftav_Tallc11} for exemplatory simulation runs the temperature
$T=\frac{2 {\overline{E}}_{\rm kin}}{3 N}$ as a function of time where
${\overline{E}}_{\rm kin}$ is the time averaged kinetic energy
with fluctuations as indicated with error bars.
We find that even after switching off the heat bath \footnote{Both the duration
of the NVT-simulation run, as well as the Nos\'e-Hoover
parameter $Q$ were chosen carefully such that $\Upsilon$ and $E_{\rm tot}$
were constant during the NVT and NVE runs respectively.}
there is no temperature drift for $T=3250\,{\rm K}$ and $T=3000\,{\rm K}$
and for $T=2500\,{\rm K}$ and $T=2750\,{\rm K}$
there is only a slight temperature drift which is of the same order as the 
temperature fluctuations and the drift occurs only for $t \lessapprox 0.6$ ns. 
For all times $t \gtrapprox 0.6$ ns and for all investigated temperatures 
there is no temperature drift and thus the comparison with previous 
simulations valid.

%
\section{Results}
In all following, we investigate how the structure and dynamics of the
system depend on the waiting time $t_{\rm w}$ elapsed after the quench
from $T_{\rm i}$ to $T_{\rm f}$. We varied the waiting time in the
range $0$\,ns $\le t_{\rm w} \le 23.98$\,ns.

\subsection{Partial Structure Factor}
\label{sec:sofq}
\begin{figure}
\includegraphics[width=3.4in]{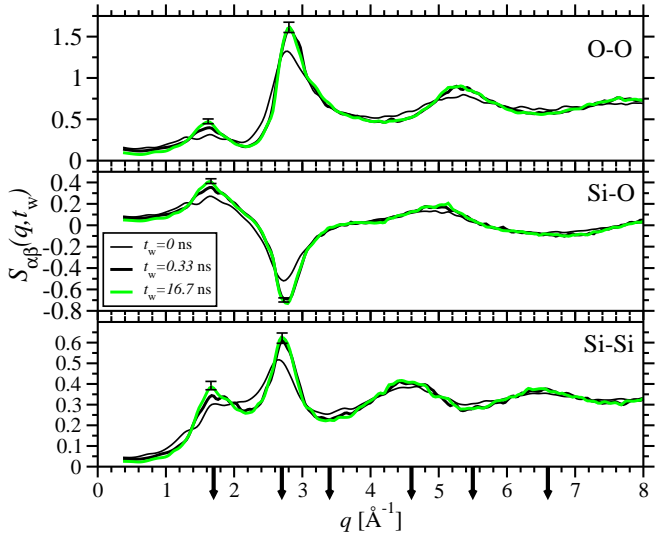}
\caption{\sf  
(Color online) Partial structure factors $S_{\alpha \beta}(q,t_{\rm w})$ as defined 
in Eq.~(\ref{eq:sofq}) for the temperature quench 
$T_{\rm i}=5000$\,K to $T_{\rm f}=2500$\,K. Indicated with arrows 
are the wave vectors $q$ which have been used to determine
$C_q(t_{\rm w},t_{\rm w}+t)$ as defined in Eq.~(\ref{eq:Cqoftwt}).}
\label{sofq_2500from5000_v4}
\end{figure}

Figure \ref{sofq_2500from5000_v4} shows for the temperature quench
$T_{\rm i}=5000$ K to $T_{\rm f}=2500$ K the partial structure factors
\cite{glassbook}
\begin{equation}
\label{eq:sofq}
S_{\alpha \beta}(q,t_{\rm w})=
\frac{1}{N} \left \langle \sum \limits_{i=1}^{N_{\alpha}}
 \sum \limits_{j=1}^{N_{\beta}} {\rm e}^{i {\bf q} \cdot
      \big ({\bf r}_i(t_{\rm w})-{\bf r}_j(t_{\rm w})\big)} 
\right \rangle
\end{equation}
where ${\bf r}_i$ and ${\bf r}_j$ are the positions of particles $i$
and $j$ of species $\alpha, \beta = {\rm O, Si}$.  The partial structure
factors for all other $(T_{\rm i}, T_{\rm f})$ combinations are very
similar.  Although Fig.~\ref{sofq_2500from5000_v4} shows $S(q,t_{\rm w})$
for the largest investigated temperature quench, there is only a slight
$t_{\rm w}$-dependence for very short waiting times $t_{\rm w} \le 0.33$\,ns 
and almost no $t_{\rm w}$-dependence for $t_{\rm w} \ge 0.33$\,ns.

%
\subsection{Generalized Incoherent Intermediate Scattering Function}
\label{sec:Cq}
\begin{figure}
\includegraphics[width=3.4in]{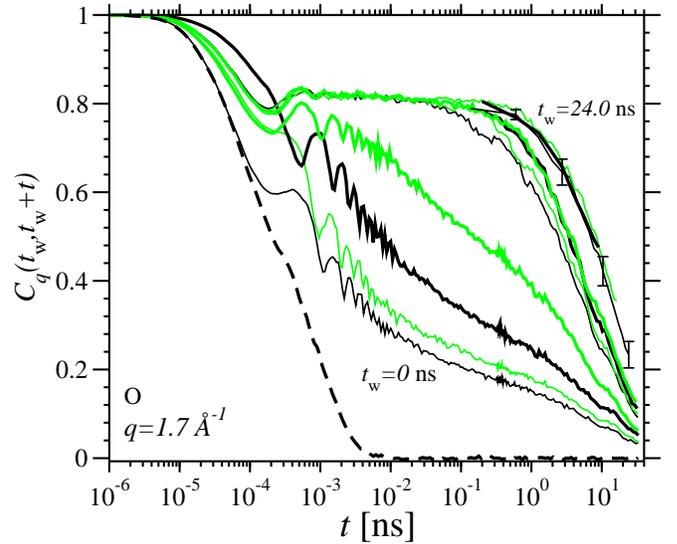}
\caption{\sf 
(Color online) $C_q(t_{\rm w},t_{\rm w}+t)$ as defined in Eq.~(\ref{eq:Cqoftwt}) for
the quench $T_{\rm i}=5000$\,K to $T_{\rm f}=2500$\,K for O-atoms and
$q=1.7$\,\AA$^{-1}$.  Waiting times for the solid lines are from bottom to
top $t_{\rm w}=0$\,ns, $1.63 \cdot 10^{-4}$\,ns, $1.63 \cdot 10^{-3}$\,ns,
$1.63 \cdot 10^{-2}$\,ns, $0.33$\,ns, $0.49$\,ns, $1.17$\,ns, $1.96$\,ns,
$8.83$\,ns, $16.67$\,ns, and $23.98$\,ns.  The order of solid lines
are for increasing $t_{\rm w}$ black thin, green (gray) thin, black thick, 
green (gray)
thick, black thin, green (gray) thin, etc.  Error bars are as indicated exemplary.
The thick dashed line corresponds to $C_q(t_{\rm w}, t_{\rm w}+t)=F_{\rm
s}(q,t)$ at $5000$\,K.}
\label{fsqt_2500from5000_q17_O_v4}
\end{figure}

\begin{figure}
\includegraphics[width=3.4in]{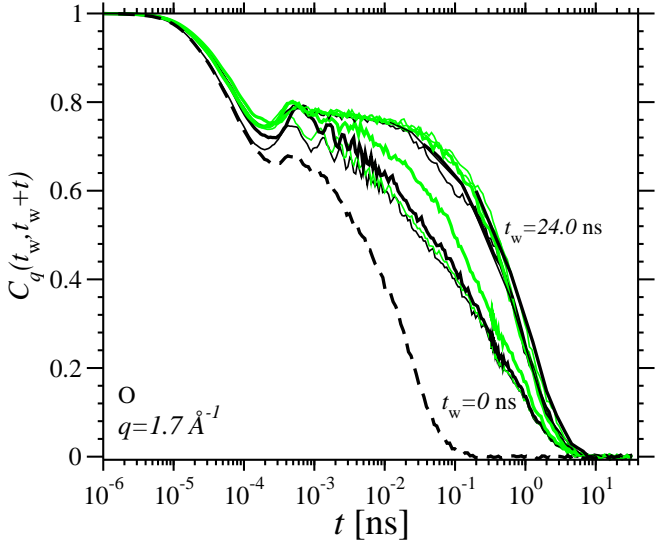}
\caption{\sf
(Color online) $C_q(t_{\rm w},t_{\rm w}+t)$ for the quench $T_{\rm i}=3760$\,K to
$T_{\rm f}=3000$\,K.  Waiting times and corresponding line styles are
the same as in Fig.~\ref{fsqt_2500from5000_q17_O_v4}.  Error bars are
of the same order as indicated in Fig.~\ref{fsqt_2500from5000_q17_O_v4}.
The thick dashed line corresponds to $C_q(t_{\rm w}, t_{\rm w}+t)=F_{\rm
s}(q,t)$ at $3760$\,K.}
\label{fsqt_3000from3760_q17_O_v4}
\end{figure}

\begin{figure}
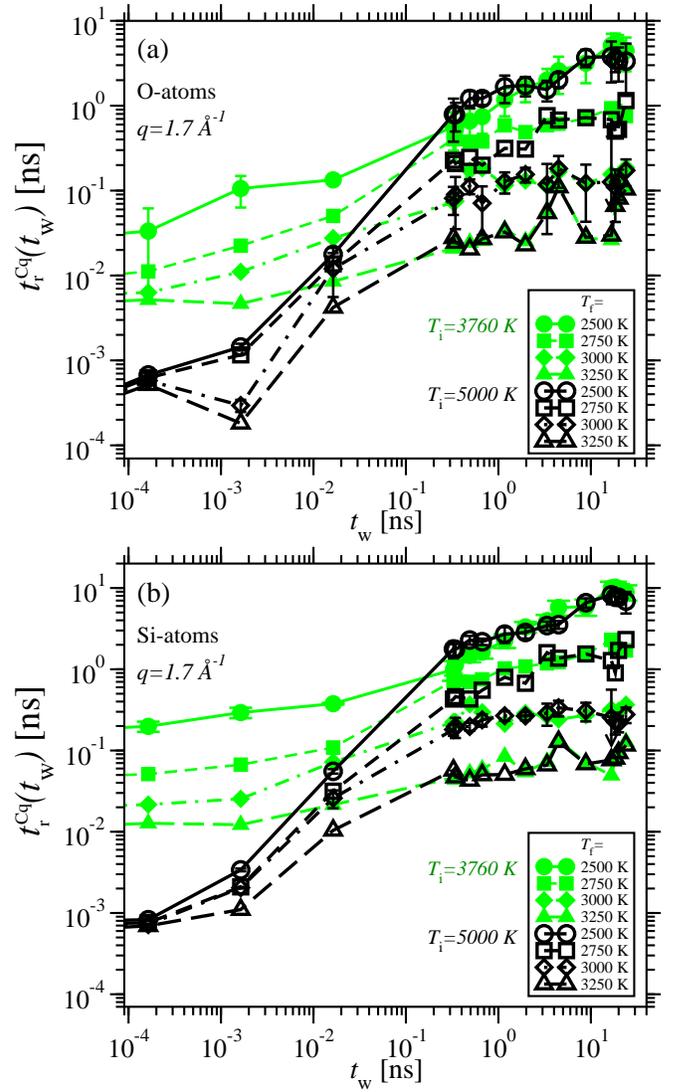

\includegraphics[width=3.4in]{fig7a.eps}
\includegraphics[width=3.4in]{fig7b.eps}
\caption{\sf 
(Color online) Decay time $t_{\rm r}^{\rm Cq}$ for $q=1.7$\,\AA$^{-1}$ and (a) O-atoms
and (b) Si-atoms.  Green (gray) is used for $T_{\rm i}=3760$\,K and black for
$T_{\rm i}=5000$\,K. Symbol shape indicates $T_{\rm f}$ as given in
the legend.  Error bars are given exemplary for ($T_{\rm i}=3760$\,K,
$T_{\rm f}=2500$\,K), ($T_{\rm i}=5000$\,K, $T_{\rm f}=2500$\,K) and
($T_{\rm i}=5000$\,K, $T_{\rm f}=3000$\,K).}
\label{tau_O_q17_v4}
\end{figure}

\begin{figure}
\includegraphics[width=3.4in]{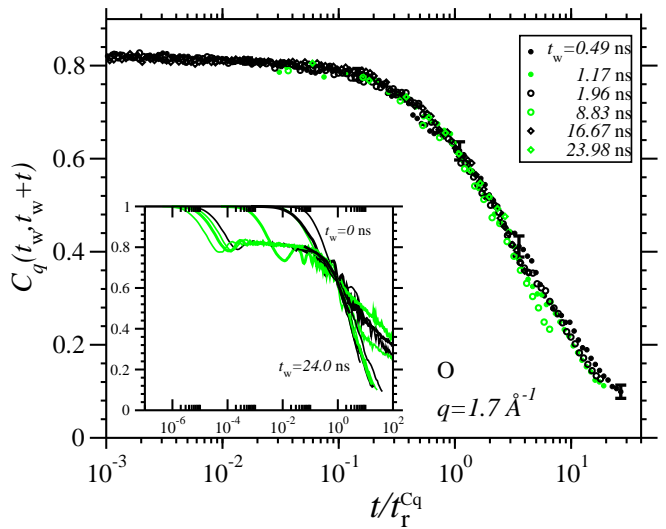}
\caption{\sf 
(Color online) $C_q(t/t_{\rm r}^{\rm Cq})$ for the quench from $T_{\rm i}=2500$\,K to
$T_{\rm i}=5000$\,K for O-atoms and $q=1.7$\,\AA$^{-1}$.  
Waiting times and corresponding lines in the inset are the same as in
Fig.~\ref{fsqt_2500from5000_q17_O_v4}.}
\label{fsqtovertau_nv_2500from5000_q17_O_v5}
\end{figure}

In this section we focus on the time-dependent generalized intermediate
incoherent scattering function \cite{glassbook}
\begin{equation}
\label{eq:Cqoftwt}
C_q(t_{\rm w},t_{\rm w}+t)=\frac{1}{N_{\alpha}}
   \left \langle \sum \limits_{j=1}^{N_{\alpha}}
      {\rm e}^{i {\bf q} \cdot
  \big ({\bf r}_j(t_{\rm w}+t)-{\bf r}_j(t_{\rm w})\big )} \right \rangle
\end{equation}
which is a measure for the correlations of the positions at time $t_{\rm
w}$ and at a later time $(t_{\rm w}+t)$.  We investigated wave vectors
of magnitude $q = 1.7, 2.7, 3.4, 4.6, 5.5$ and $6.6$\,\AA$^{-1}$,
as indicated with arrows in Fig.~\ref{sofq_2500from5000_v4}.
We show in Fig.~\ref{fsqt_2500from5000_q17_O_v4}
and Figs.~\ref{fsqt_3000from3760_q17_O_v4}
- \ref{fsqtovertau_nv_2500from5000_q17_O_v5} results for the
first sharp diffraction peak at $q=1.7$\,\AA$^{-1}$. Similar
results are found for all other investigated wave vectors.
Figure~\ref{fsqt_2500from5000_q17_O_v4} shows $C_q(t_{\rm w},t_{\rm
w}+t)$ for the largest investigated temperature quench from $T_{\rm
i}=5000$\,K to $T_{\rm f}=2500$\,K for waiting times $t_{\rm w}=0$
- 23.98\,ns, as listed in the figure caption of 
Fig.~\ref{fsqt_2500from5000_q17_O_v4}.
We find that $C_q(t_{\rm w},t_{\rm w}+t)$ is dependent on $t_{\rm w}$
for all but the last three investigated waiting times.  For very short
times $t \lesssim 5 \cdot 10^{-5}$ ns and zero waiting time, $C_q(t_{\rm
w}=0,t)$ is well approximated by $C_q$ of the high temperature $T_{\rm
i}=5000$ K from which the system has been quenched (see dashed line in
Fig.~\ref{fsqt_2500from5000_q17_O_v4}).  Thus, $C_q(t_{\rm w}=0,t)$ for
very short times is only dependent on $T_{\rm i}$, $q$ and the particle
type, but independent of $T_{\rm f}$.  For times of the order of $t =
10^{-3}$\,ns, $C_q(t_{\rm w},t_{\rm w}+t)$ is oscillatory due to the small
system size. For times $t \gtrsim 10^{-3}$\,ns, $C_q$ decays to zero
without forming a plateau for small $t_{\rm w}$.  With increasing $t_{\rm
w}$ a plateau is formed, which is independent of $t_w$ for $t_{\rm w}
\ge 0.33$\,ns.

\begin{figure}
\includegraphics[width=3.4in]{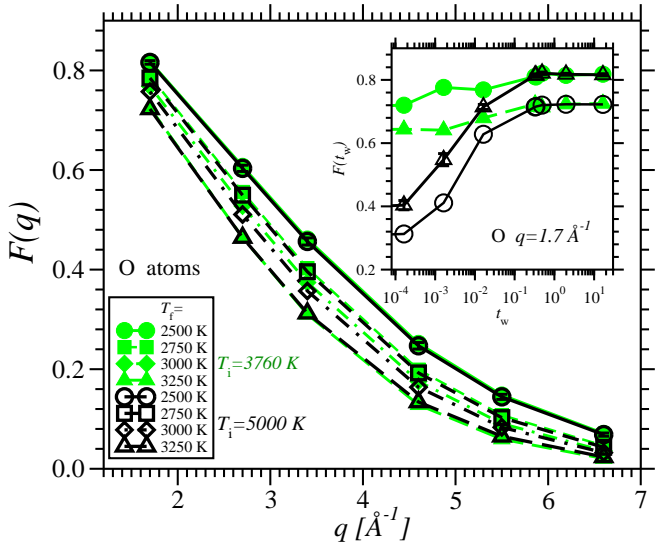}
\caption{\sf
(Color online) Plateau height, $F$, as defined in the text, as a function of wave vector
$q$ ($t_{\rm w}=16.67$\,ns) and in the inset as a function of $t_{\rm w}$
($q=1.7$\,\AA$^{-1}$).  Error bars are of the same size as symbols as
indicated exemplary for $T_{\rm i}=5000$ K and $T_{\rm f}=2500$ K.}
\label{Fofq_O_tw10000000_v3}
\end{figure}

To characterize the plateau height we define $F$ as the time average
of $C_q(t_{\rm w},t_{\rm w}+t)$ for times $2.55$\,ps $\le t \le
6.64$\,ps. The inset of Fig.~\ref{Fofq_O_tw10000000_v3} shows that
for large waiting times $F(t_{\rm w})$ becomes independent of $t_{\rm
w}$ and of $T_{\rm i}$.  To test this independence of $T_{\rm i}$
further, we show $F$ as a function of $q$ for $t_{\rm w}=16.67$\,ns
in Fig.~\ref{Fofq_O_tw10000000_v3}.  We find that the plateau height
is dependent on the particle type and decreases with decreasing $q$,
but $F$ is independent of $T_{\rm i}$.


\vspace*{10mm}

The plateau in $C_q$ becomes more horizontal with decreasing
final temperature $T_{\rm f}$, as can be seen by the comparison
of Fig.~\ref{fsqt_2500from5000_q17_O_v4} ($T_{\rm f}=2500$\,K) and
Fig.~\ref{fsqt_3000from3760_q17_O_v4} ($T_{\rm f}=3000$\,K). Times $t
\gtrsim 0.1$\,ns correspond to the $\alpha$-relaxation, where $C_q(t_{\rm
w},t_{\rm w}+t)$ decays from the plateau to zero.  For the quench from
$5000$\,K to $2500$\,K (see Fig.~\ref{fsqt_2500from5000_q17_O_v4})
this decay depends on $t_{\rm w}$ for all $t_{\rm w} < 8.83$\,ns.
However, for $t_{\rm w} \ge 8.83$\,ns (the largest three $t_{\rm w}$),
$C_q(t_{\rm w},t_{\rm w}+t)$ is independent of $t_{\rm w}$ not only
for intermediate times $t$ (plateau) but for all times (including the
$\alpha$-relaxation).  Thus, the system reaches equilibrium during
the simulation run.  For the quench from $3760$\,K to $3000$\,K (see
Fig.~\ref{fsqt_3000from3760_q17_O_v4}), $C_q(t_{\rm w},t_{\rm w}+t)$
becomes independent of $t_{\rm w}$ for $t_{\rm w} \gtrsim 1$\,ns,  which
means that the time at which equilibrium is reached is dependent on the
temperature quench.

To estimate the time when the system reaches equilibrium for each $(T_{\rm
i}, T_{\rm f})$ combination, we next quantify the decay of $C_q(t_{\rm
w},t_{\rm w}+t)$ .  Instead of taking a vertical cut in $C_q(t_{\rm
w},t_{\rm w}+t)$ as we did for $F$, we now take a horizontal cut. We
define the decay time $t_{\rm r}^{\rm Cq}$ to be the time 
$t=t_{\rm r}^{\rm Cq}$ for which 
$C_q(t_{\rm w},t_{\rm w}+t_{\rm r}^{\rm Cq})=C_{\rm
cut}$.  We chose $C_{\rm cut}=0.625/0.41/0.295/0.155/0.085/0.04)$ for the
Si particles and $C_{\rm cut}=0.625/0.305/0.195/0.08/0.04/ 0.014$ for the
O particles at the wavenumbers $q=1.7/2.7/3.4/4.6/5.5/6.6$\,\AA$^{-1}$,
respectively. The resulting decay times $t_{\rm r}^{\rm Cq}$ as a
function of waiting time $t_{\rm w}$ are shown in Fig.~\ref{tau_O_q17_v4}a
for O-atoms and in Fig.~\ref{tau_O_q17_v4}b for Si-atoms.  Color (black
or green/gray) indicates the initial temperature $T_{\rm i}$ and symbol shape
indicates the final temperature $T_{\rm f}$.

We find that $t_{\rm r}^{\rm Cq}(t_{\rm w})$ is characterized by three
different $t_{\rm w}$-windows.  (I) For waiting times $t_{\rm w} \lesssim
0.3$\,ns, decay times are significantly lower for $T_{\rm i}=5000$\,K
(black lines and symbols) than for $T_{\rm i}=3760$\,K (green/gray lines
and symbols).  The dependence of $t_{\rm r}^{\rm Cq}$ on $t_{\rm w}$
is strongly dependent on all varied parameters, i.e.~$T_{\rm i}$,
$T_{\rm f}$, particle type, and $q$.  
For $T_{\rm i}=5000$\,K, $T_{\rm f}=2500$\,K, $2750$\, K, and 
$q=1.7$ \AA$^{-1}$, $2.7$ \AA$^{-1}$ 
$t_{\rm r}^{\rm Cq}(t_{\rm w})$ follows roughly a power
law with an exponent $\mu \approx 1.15$ with variations of the order
of $0.07$ dependent on $T_{\rm f}$,
particle type, and $q$. 
(II) For intermediate waiting times, $t_{\rm
r}^{\rm Cq}(t_{\rm w})$ also follows roughly a power law with a different
exponent than in regime (I). We find for $T_{\rm i}=5000$\,K, 
$T_{\rm f}=2500$\,K, $2750$\,K and $q=1.7$ \AA$^{-1}$, $2.7$ \AA$^{-1}$
$\mu \approx 0.35$ with variations of the order of $0.08$ depending 
on $T_{\rm f}$, particle type, and $q$.   
Best power law fits are for $T_{\rm i}=3760$\,K, $T_{\rm f}=2500$\,K
with $\mu$ ranging from $\mu=0.55$/$0.57$ for $q=1.7$ \AA$^{-1}$
to $\mu=0.69$/$0.63$ for $q=6.6$ \AA$^{-1}$ and for Si/O atoms.
Kob and Barrat  find for the binary Lennard-Jones 
system also a power law for $t_{\rm r}^{\rm Cq}(t_{\rm w})$, however, 
with $\mu=0.882$ \cite{KobBarratPRL78}. Similar to Grigera et al. 
\cite{grigera2004} we find that the transition from small waiting times (I)
to intermediate waiting times (II) is accompanied by a change of the
exponent $\mu$.
(III) For very long waiting times $t_{\rm
r}^{\rm Cq}(t_{\rm w})$ is independent of $t_{\rm w}$ and $T_{\rm i}$,
i.e.~equilibrium is reached.  The waiting time $t_{23}$ for which the
transition from regime (II) to regime (III) occurs is dependent on
$T_{\rm f}$: $t_{23} \approx 0.3$\,ns for $T_{\rm f}=3250$\,K, $t_{23}
\approx 1$\,ns for $T_{\rm f}=3000$\,K, $t_{23} \approx 3$\,ns for
$T_{\rm f}=2750$\,K, and $t_{23} \approx 10$\,ns 
for $T_{\rm f}=2500$\,K.\footnote{These estimates 
of $t_{23}$ are in good agreement with the results of
Scheidler et al. \cite{scheidler2001} who determined the relaxation 
time via the specific heat and who found good agreement with experimental 
data.}

Mean-field spin glass models predict \cite{MFCq51,MFCq52}
\begin{equation}
\label{eq:CqMF}
C_q(t_{\rm w},t_{\rm w}+t)=
             C_q^{\rm ST}(t) + C_q^{\rm AG}\left (
                  \frac{h(t_{\rm w}+t)}{h(t_{\rm w})} \right ) \; ,
\end{equation}
according to which $C_q(t_{\rm w},t_{\rm w}+t)$ can be separated into a
short-time term $C_q^{\rm ST}(t)$ that is independent of $t_w$ and an
intermediate-time term that 
scales as $h(t_{\rm w}+t)/h(t_{\rm w})$ where $h$ is a monotonously
increasing function.
%
%
%
It has been observed for different systems that the function $h(t)$
follows $h(t) \approx t^{\alpha}$. Thus, the so-called ``simple aging''
(see \cite{KobBarratPRL78} and references therein) applies and, as
a consequence, $C_q$ as a function of $(t/t_{\rm w})$ for different
$t_{\rm w}$ superimpose.

M{\"u}ssel and Rieger \cite{CqRieger} have proposed activated dynamics
for $C_q(t_{\rm w}, t_{\rm w}+t)$,
\begin{equation}
\label{eq:CqRieger}
C_q(t_{\rm w},t_{\rm w}+t)=
    C_q^{\rm ST}(t) + C_q^{\rm AG}\left (
           \frac{{\rm ln}
               \left [(t_{\rm w}+t)/\tau_{\rm fit} \right ] }
             {{\rm ln} \left [t_{\rm w}/\tau_{\rm fit} \right ] } 
           \right ),
\end{equation}
where the characteristic time scale $\tau_{\rm fit}$ is a fit parameter.
We find that neither $C_q \left ( \frac{t_{\rm w}+t}{t_{\rm w}} \right
)$, nor $C_q \left ( \frac{t}{t_{\rm w}} \right )$, nor $C_q \left (
\frac{{\rm ln} \left [(t_{\rm w}+t)/\tau_{\rm fit} \right ]} {{\rm ln}
\left [t_{\rm w}/\tau_{\rm fit} \right ]} \right )$ (for any choice of
$\tau_{\rm fit}$) superimpose for different $t_{\rm w}$.  Instead we
find, similar to the results of Kob and Barrat \cite{KobBarratPRL78} for
a binary Lennard-Jones system, that time superposition holds, defined by
\begin{equation}
\label{eq:CqSimpleAging}
C_q(t_{\rm w},t_{\rm w}+t)=
    C_q^{\rm ST}(t) + C_q^{\rm AG}\left (
   \frac{t}{t_{\rm r}^{\rm Cq}(t_{\rm w})} \right ) \, .
\end{equation}
In Fig.~\ref{fsqtovertau_nv_2500from5000_q17_O_v5}, we show
$C_q(t/t_{\rm r}^{\rm Cq})$ for the same set of parameters as in
Fig.~\ref{fsqt_2500from5000_q17_O_v4}. When all waiting times are
included (see inset of Fig.~\ref{fsqtovertau_nv_2500from5000_q17_O_v5})
time superposition does not apply due to including too short waiting
times.  Wahlen and Rieger \cite{WahlenRieger} have studied $C_q(t_{\rm
w},t_{\rm w}+t)$ for the same BKS-SiO$_2$ system, however for waiting
times smaller than $50$\,ps. Their results are consistent with
the inset of Fig.~\ref{fsqtovertau_nv_2500from5000_q17_O_v5}.
For waiting times $t_{\rm w} \ge 0.49$\,ns (see
Fig.~\ref{fsqtovertau_nv_2500from5000_q17_O_v5}), however,
Eq.~(\ref{eq:CqSimpleAging}) is a good approximation.  Please note that
$C_q(t/t_{\rm r}^{\rm Cq})$ follows time superposition for all waiting
times $t_{\rm w} \ge 0.49$\,ns, i.e.~not only for the time-range (II), but
also for the time-range (III). That means for the $\alpha$-relaxation
that the shape of the out-of equilibrium curves is the same (within error
bars) as the shape of the equilibrium curves.  We find similar results
for all other $(T_{\rm i},T_{\rm f})$ combinations, Si-particles, and
all other $q$.

\begin{figure}
\includegraphics[width=3.4in]{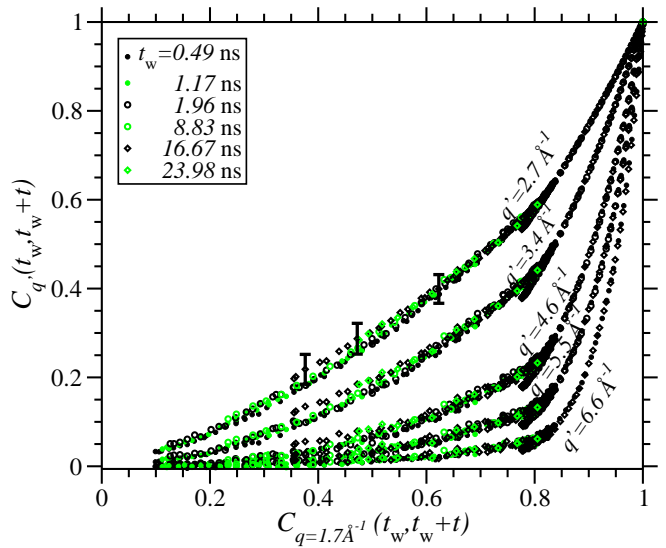}
\caption{\sf 
(Color online) $C_{q'}(t_{\rm w},t_{\rm w}+t)$ as a function of $C_q(t_{\rm
w},t_{\rm w}+t)$ for $q=1.7$\,\AA$^{-1}$ and $q'=2.7$, 3.4, 4.6, 5.5,
6.6\,\AA$^{-1}$ for O-atoms and the temperature quench from $5000$\,K
to $2500$\,K.}
\label{fq17f2_2500from5000_nveB_v3}
\end{figure}

Next we test whether 
$C_q=C(q,t_{\rm w},t_{\rm w}+t)$
scales as $C_q=C(q,z(t_{\rm w},t)$ where $z$ is a function of $t_{\rm w}$ and
$t$ only, i.e.~independent of $q$.  
Following an approach of Kob and Barrat
\cite{KobBarratCq1Cq2} we show in  Figure~\ref{fq17f2_2500from5000_nveB_v3} 
a parametric plot for $C_{q'}(t_{\rm w},t_{\rm w}+t)$ as a function of
$C_q(t_{\rm w},t_{\rm w}+t)$ for $q=1.7$\,\AA$^{-1}$ and $q'=2.7$, 3.4,
4.6, 5.5, 6.6\,\AA$^{-1}$ for O-atoms and the temperature quench from
$5000$\,K to $2500$\,K.  For sufficiently large $t_{\rm w}$ we find,
contrary to the results of Kob and Barrat for the Lennard-Jones system,
that for SiO$_2$ the parametric curves superimpose and thus that 
$C(q,t_{\rm w},t_{\rm w}+t)=C(q,z(t_{\rm w},t)$ 
for $t_{\rm w} \ge 0.49$ ns.  This includes,
within error bars, also the equilibrated curves for $t_{\rm w} \gtrsim
10$\,ns. We find similar results for all other $(T_{\rm i},T_{\rm f})$
combinations, Si-particles, and all other $q$.

%
\subsection{Mean Square Displacement}
\label{sec:msd}

In the previous section, we have focused on the analysis of $C_q(t_{\rm
w},t_{\rm w}+t)$ and identified different time-windows. In this section,
we consider the mean square displacement
\begin{equation}
\label{eq:msd}
\Delta r^2 (t_{\rm w},t_{\rm w}+t) \, = \,
 \frac{1}{N} \, \sum \limits_{i=1}^{N}
 \left \langle
 \left ( {\bf r}_i(t_{\rm w}+t) - {\bf r}_i(t_{\rm w}) \right )^2 
 \right \rangle
\, .
\end{equation}

\begin{figure}
\includegraphics[width=3.4in]{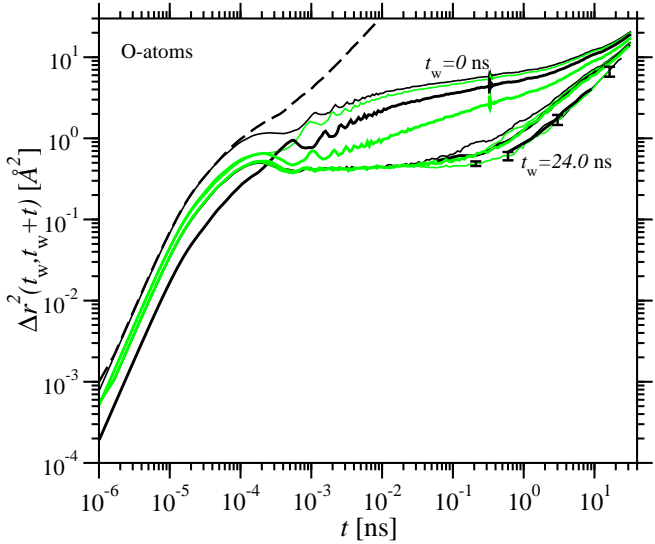}
\caption{\sf  
(Color online) Mean square displacement $\Delta r^2 (t_{\rm w},t_{\rm w}+t)$ as defined
in Eq.~(\ref{eq:msd}) for the temperature quench from $5000$\,K to
$2500$\,K and for O-atoms.
Waiting times and corresponding line styles are the same as in
Fig.~\ref{fsqt_2500from5000_q17_O_v4}.}
\label{msdoft_2500from5000_nv_O_v3}
\end{figure}

Figure~\ref{msdoft_2500from5000_nv_O_v3} shows $\Delta r^2(t_{\rm
w},t_{\rm w}+t)$ for the temperature quench from $5000$\,K to $2500$\,K
and for O-atoms.  As in Fig.~\ref{fsqt_2500from5000_q17_O_v4}, for times
$t \lesssim 5 \cdot 10^{-5}$\,ns and zero waiting time, the mean square
displacement $\Delta r^2(t_{\rm w}=0,t)$ is well approximated by $\Delta
r^2$ of the high temperature $T_{\rm i}=5000$\,K from which the system has
been quenched (see dashed line in Fig.~\ref{msdoft_2500from5000_nv_O_v3})
and thus independent of $T_{\rm f}$.  For times $t \approx 10^{-3}$\,ns,
$\Delta r^2(t_{\rm w},t_{\rm w}+t)$ is oscillatory due to the small
system size \cite{JuergenFSFs}, while for times $t \gtrsim 10^{-3}$\,ns
and waiting times $t_{\rm w} \ge 0.33$\,ns, we find that $\Delta r^2$
forms a plateau which is independent of $t_w$.  As for $C_q$, we find that
the plateau is the more horizontal the smaller $T_{\rm f}$ and the plateau
height depends on the particle type but is independent of $T_{\rm i}$.

\begin{figure}
\includegraphics[width=3.4in]{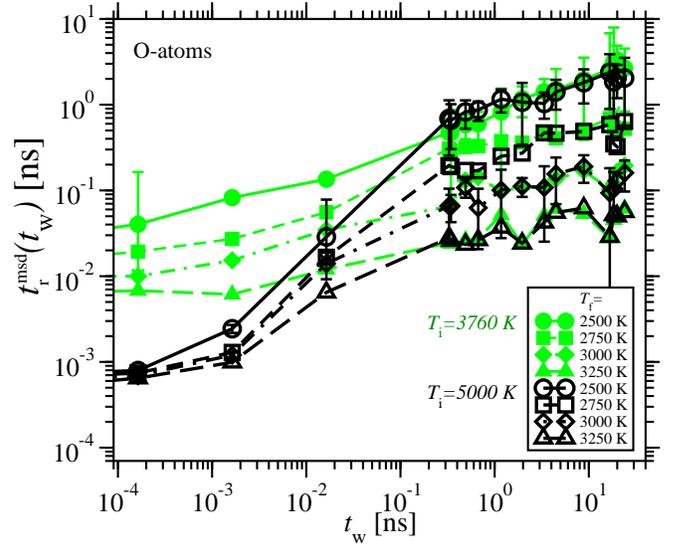}
\caption{\sf 
(Color online) $t_{\rm r}^{\rm msd}$ for O-atoms.  Symbols for the different $(T_{\rm
i},T_{\rm f})$ combinations are the same as in Fig.~\ref{tau_O_q17_v4}.
Error bars are indicated exemplary for ($3760$\,K, $2500$\,K), ($5000$\,K,
$2500$\,K) and ($5000$\,K, $3000$\,K).}
\label{taumsd_v2_O}
\end{figure}

For waiting times $t_{\rm w} \ge 0.33$\,ns and times $t \gtrsim 0.1$\,ns,
the mean square displacement leaves the plateau and increases further. To
characterize the dependence of this $\alpha$-relaxation we define the
time $t_{\rm r}^{\rm msd}$ as the time $t=t_{\rm r}^{\rm msd}$ for which
$\Delta r^2(t_{\rm w},t_{\rm w}+t_{\rm r}^{\rm msd})=1.35$\,\AA$^2$ (see
Fig.~\ref{taumsd_v2_O}).  We can identify again the three time windows
(I) of waiting times $t_{\rm w} \lesssim 0.3$\,ns with a dependence on
$T_{\rm i}$, $T_{\rm f}$ and particle type, (II) the aging regime of
intermediate waiting times where $t_{\rm r}^{\rm msd}$ follows roughly
a power law, and (III) for very long waiting times when equilibrium
is reached.  The transition from (II) to (III) occurs at approximately
the same times $t_{23}$ as for $C_q$, i.e.~$t_{23} \approx 0.3$\,ns for
$T_{\rm f}=3250$\,K, $t_{23} \approx 1$\,ns for $T_{\rm f}=3000$\,K,
$t_{23} \approx 3$\,ns for $T_{\rm f}=2750$\,K and $t_{23} \approx
10$\,ns for $T_{\rm f}=2500$\,K.

\begin{figure}
\includegraphics[width=3.4in]{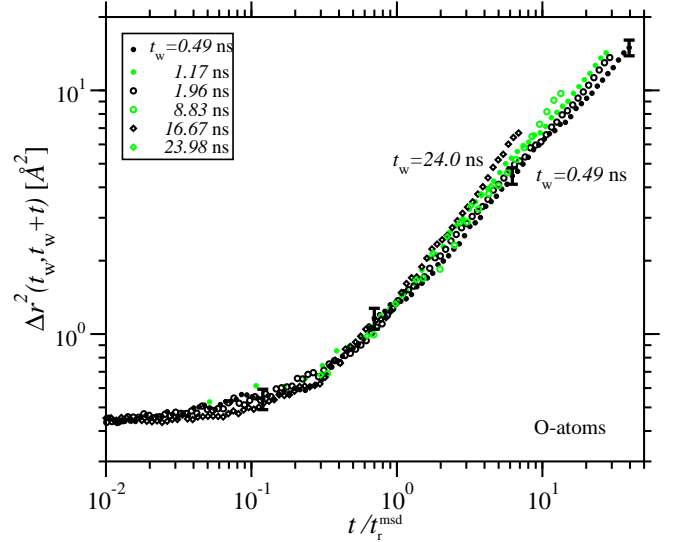}
\caption{\sf 
(Color online) $\Delta r^2(t/t_{\rm r}^{\rm msd})$ for the temperature quench from
$5000$\,K to $2500$\,K and for O-atoms.}
\label{msdoftovertaumsd_2500from5000_nv_O_v3}
\end{figure}

Figure~\ref{msdoftovertaumsd_2500from5000_nv_O_v3} shows the equivalent
of Fig.~\ref{fsqtovertau_nv_2500from5000_q17_O_v5} to test time
superposition. We find for $\Delta r^2(t/t_{\rm r}^{\rm msd})$ that
time superposition is valid for waiting times $0.34$\,ns $\le t_{\rm w}
\lesssim 8.83$\,ns, i.e.~for the time window (II) but not for the time
window (III).

%
\section{Summary}

Using molecular dynamics simulations, we investigated for the strong
glass former SiO$_2$ the aging dynamics below the critical MCT temperature
$T_c$, using the BKS potential to model the interactions between silicon
and oxygen atoms.  After an instantaneous quench from $T_{\rm i}>T_c$ to a
temperature $T_{\rm f}<T_c$ the dynamics towards equilibrium was studied
as a function of waiting time $t_{\rm w}$.  Note that the temperatures
$T_{\rm f}$ were chosen such that equilibrium was reached on the time
span of the simulations (of the order of 30\,ns).  The central quantities
considered in this work are the incoherent intermediate scattering function
$C_q(t_{\rm w}, t_{\rm w}+t)$ and the mean square displacement 
$\Delta r^2(t_{\rm w}, t_{\rm w}+t)$.
These functions depend on the time origin at
$t_{\rm w}$ as long as $t_{\rm w}$ is smaller than the typical relaxation
time, $\tau_{\rm eq}$, that is required to equilibrate the system.

We find that the decay of $C_q(t_{\rm w}, t_{\rm w}+t)$ 
(and similarly the rise of $\Delta r^2(t_{\rm w}, t_{\rm w}+t)$) exhibit
qualitative changes from short to long waiting times. At short waiting
times, relaxation processes are dominant that correspond to the early $\beta$
relaxation regime at the target temperature $T_{\rm f}$. In this
$t_{\rm w}$ regime, no well-defined plateau is found in $C_q(t_{\rm w},
t_{\rm w}+t)$ (see Fig.~\ref{fsqt_2500from5000_q17_O_v4}).  Instead,
this function first decreases rapidly, followed by a strongly stretched
exponential decay to zero.  At long waiting times, the $\beta$ relaxation
seems to be very similar to that at equilibrium. The Debye-Waller factor
(i.e.~the height of the plateau in $C_q(t_{\rm w}, t_{\rm w}+t)$) has
reached its equilibrium value, although the decay of $C_q(t_{\rm w},
t_{\rm w}+t)$ from the plateau to zero is faster than that at equilibrium.
However, the shape of curves describing the long-time decay of $C_q(t_{\rm
w}, t_{\rm w}+t)$ is the same as that at equilibrium. Thus, $C_q$ follows
a simple time superposition for long waiting times, different e.g.~from
the ``activated dynamics scaling'' proposed by Wahlen and Rieger for
BKS silica.

Our results show that the aging dynamics of BKS silica is very similar
to that of the KALJ mixture.  For both silica and the KALJ mixture
three $t_{\rm w}$-regimes can be identified and $C_q$ follows time
superposition for sufficiently large $t_{\rm w}$. The only difference
between these two systems is that $C_q$ scales as $C(q,z(t_{\rm w},t)$
for SiO$_2$ but less well for the KALJ mixture.  So slightly below
its critical temperature $T_{\rm c}$ of MCT, the strong glass-former
silica does not seem to be very different from typical fragile systems,
although one has already reached the low temperature Arrhenius regime
(note that activation energies for the self-diffusion, viscosity etc. are
of the order of 5\,eV, similar to the corresponding activation energies
close to the glass transition temperature $T_{\rm g}\approx 1450$\,K,
as measured in various experiments).  However, the dynamics could be
very different at very low temperatures (close to $T_{\rm g}$) where the
long-time aging regime is not accessible by computer simulations. Thus,
more experimental work on the aging dynamics of silica around $T_{\rm g}$
would be very desirable.  We also leave for future work to test whether
also for other systems three $t_{\rm w}$-regimes are identified and
whether the equilibrium curve is included in the time superposition of
$C_q$ at intermediate and large $t_{\rm w}$.

\begin{acknowledgments}
KVL thanks A. Zippelius and the Institute of Theoretical Physics,
University of G\"ottingen, for hospitality and financial support.
\end{acknowledgments}

\appendix

\section{Velocity Verlet Nos\'e Hoover}
\label{NoseHooverEqu} 
  The Nos\'e Hoover equations of motion are for particles $i=1,\ldots,N$
  at position $\bf{r}_i$ with momentum $\bf{p}_i$ 
  \begin{eqnarray}
    \dot{\bf r}_i & = & \frac{{\bf p}_i}{m_i} \\
    \dot{\bf p}_i & = & {\bf F}_i - \zeta {\bf p}_i \\
    \dot{\zeta} & = &\frac{1}{Q} 
        \left ( \sum \limits_{i=1}^{N} \frac{{\bf p}_i^2}{m_i} - X k T
        \right )
  \end{eqnarray}
and thus
  \begin{eqnarray}
    \ddot{\bf r}_i & = &  \frac{{\bf F}_i}{m_i} - \zeta \, \dot{\bf r}_i \\
    \frac{{\rm d}^2 \ln s}{{\rm d}t^2}  =  \dot{\zeta} 
         & =  & \frac{1}{Q} 
                \left ( \sum \limits_{i=1}^{N} m_i \dot{{\bf r}}_i^2 - X k T
                \right )
  \end{eqnarray}
   where $X=3N$ .
We integrated with the generalized velocity Verlet form of
Fox and Andersen \cite{fox} 
  \begin{eqnarray}
    {\bf r}_i(t+\Delta t) & = & {\bf r}_i(t) 
          + \Delta t  \, \dot{\bf r}_i(t)  
          \\ & + &
           \frac{(\Delta t)^2}{2} 
             \left ( \frac{{\bf F}_i(t)}{m_i} 
                         - \zeta (t) \,\dot{\bf r}_i(t) \right )
          \nonumber  \\
    \ln s(t+\Delta t) & = & \ln s(t) + \Delta t \, \zeta(t) 
          \\ & + &
           \frac{(\Delta t)^2}{2 Q} 
                \left ( \sum \limits_{i=1}^{N} m_i \dot{\bf{r}}_i^2(t) - X k T
                \right ) \nonumber  \\
    \zeta^{\rm approx}(t+\Delta t) & = & \zeta(t) 
          \\ & + & \frac{\Delta t}{Q}
             \left ( \sum \limits_{i=1}^{N} m_i \dot{\bf{r}}_i^2(t) - X k T
                \right ) \nonumber   \\
   \dot{\bf r}_i(t+\Delta t) & = & \dot{\bf r}_i(t)  \\
                     & + & \frac{\Delta t}{2} 
         \bigg ( \frac{{\bf F}_i(t)+{\bf F}_i(t+\Delta t)}{m_i} 
        \nonumber   \\ &  & -  
  \big [ \zeta(t) 
                   + \zeta^{\rm approx}(t+\Delta t) \big ] 
                          \dot{\bf r}_i(t) 
          \bigg ) 
         \nonumber  \\ &  & 
       \bigg (
          1 - \frac{\Delta t}{2} \zeta^{\rm approx}(t+\Delta t) \bigg )
       \nonumber \\
   \zeta(t+\Delta t) & = & \zeta(t) + \frac{\Delta t}{2Q}  \Big (
                \sum \limits_{i=1}^{N} m_i \dot{\bf{r}}_i^2(t)   
           \\ &   & +
   \sum \limits_{i=1}^{N} m_i \dot{\bf{r}}_i^2(t+\Delta t) 
              - 2 X k T                       \Big )\nonumber 
  \end{eqnarray}
To ensure that 
$\Upsilon = 
   \sum \limits_{i=1}^{N} \frac{{\bf p}_i^2}{2 m_i} + U(\{{\bf r}_i \} )
   + \frac{Q}{2} \zeta^2 + X k T \ln s$ is conserved 
   (see \cite{branka}) we chose
    $Q=50000$ \AA$^2$ u. 

\bibliography{SiO2aging_v14}

\end{document}